# Reverse time-to-death as time-scale in time-to-event analysis for studies of advanced illness and palliative care

Yin Bun Cheung,[1,2,3] * Xiangmei Ma,[2] Isha Chaudhry,[1,4] Nan Liu,[1,2,5] Qingyuan Zhuang,[6,7] Grace Meijuan Yang,[1,4,6] Chetna Malhotra,[1,4] Eric Andrew Finkelstein [1,4]

1. Programme in Health Services & Systems Research, Duke-NUS Medical School, National University of Singapore, 8 College Road, Singapore 169857
2. Centre for Quantitative Medicine, Duke-NUS Medical School, National University of Singapore, 8 College Road, Singapore 169857
3. Tampere Center for Child, Adolescent and Maternal Health Research, Tampere University, Arvo Ylpön katu 34, Tampere 33520, Finland
4. Lien Center for Palliative Care, Duke-NUS Medical School, National University of Singapore, 8 College Road, Singapore 169857
5. Duke-NUS AI + Medical Science Initiative, Duke-NUS Medical School, National University of Singapore, 8 College Road, Singapore 169857
6. Division of Supportive and Palliative Care, National Cancer Centre Singapore, 30 Hospital Boulevard, Singapore 168583.
7. Data and Computational Science Core, National Cancer Centre Singapore, 30 Hospital Boulevard, Singapore 168583.

* Corresponding author:

Professor Yin Bun Cheung, Duke-NUS Medical School, 8 College Road, Singapore 169857

Email: yinbun.cheung@duke-nus.edu.sg

ORCID: 0000-0003-0517-7625

Word count: 2943

**Abstract**

**Background:** Incidence of adverse outcome events rises as patients with advanced illness approach end-of-life. Exposures that tend to occur near end-of-life, e.g., use of wheelchair, oxygen therapy and palliative care, may therefore be found associated with the incidence of the adverse outcomes. We propose a strategy for time-to-event analysis to mitigate the time-varying confounding.

**Methods:** We propose a concept of reverse time-to-death (rTTD) and its use for the time-scale in time-to-event analysis. We used data on community-based palliative care uptake (exposure) and emergency department visits (outcome) among patients with advanced cancer in Singapore to illustrate. We compare the results against that of the common practice of using time-on-study (TOS) as time-scale.

**Results:** Graphical analysis demonstrated that cancer patients receiving palliative care had higher rate of emergency department visits than non-recipients mainly because they were closer to end-of-life, and that rTTD analysis made comparison between patients at the same time-to-death. Analysis of emergency department visits in relation to palliative care using TOS time-scale showed significant increase in hazard ratio estimate when observed time-varying covariates were omitted from statistical adjustment (change-in-estimate=0.38; 95% CI 0.15 to 0.60). There was no such change in otherwise the same analysis using rTTD (change-in-estimate=0.04; 95% CI -0.02 to 0.11), demonstrating the ability of rTTD time-scale to mitigate confounding that intensifies in relation to time-to-death.

**Conclusion:** Use of rTTD as time-scale in time-to-event analysis provides a simple and robust approach to control time-varying confounding in studies of advanced illness, even if the confounders are unmeasured.

**KEY MESSAGES**

**What is already known on this topic**

Appropriate choice of time-scale for Cox model and its extensions can control time-varying confounding.

In studies of advanced illness, potentially beneficial exposures may appear to have an adverse impact on health outcomes because both the exposures and negative health outcomes tend to occur near end-of-life.

**What this study adds**

Reverse time-to-death as time-scale in time-to-event analysis is a simple and robust method for mitigating time-varying confounding in studies of advanced illness, even if the confounders are unmeasured.

**How this study may affect research, practice or policy**

Researchers applying Cox-type models for time-to-event analysis of patients with advanced illness should consider reverse time-to-death as a major candidate for the time-scale.

**INTRODUCTION**

Advanced illness imposes substantial suffering on patients and healthcare costs on society, and the burdens are projected to increase rapidly in the next few decades.[1] Palliative care aims to improve quality of life of patients and their families who are facing challenges arising from advanced illness. It is hypothesized that palliative care can also reduce acute healthcare utilization that is not effective in promoting well-being and therefore reduce healthcare costs.[2,3] However, there has been limited evidence about such benefits.[2,4] It was suggested that the differences in palliative care delivery in trial setting and real-world setting led to under-estimation of the effect of palliative care by randomized trials.[5] Observational studies, possibly using real-world data, may play an important role in the evaluation.

Multivariable regression and propensity score methods have been used in a multitude of observational studies of emergency department (ED) visits, hospital admissions and hospital cost, including in the palliative care setting.[6] However, both methods rely on a critical assumption of no unobserved confounders, which is difficult to ascertain.[7] A recent systematic review maintained that while there "exists a large volume of studies using multivariable regression or propensity score approaches to control for observed confounding, … there has been insufficient attention paid to unobserved confounding and selection bias."[6] An alternative approach that may handle unobserved confounding is difference-in-difference analysis, which compares rates of change in outcomes over time between populations that do and do not experience introduction of a policy intervention.[8,9]

In the setting of advanced illness, studies of exposures that tend to occur near end-of-life, such as use of wheelchair, oxygen therapy and palliative care, may suffer a high level of time-varying confounding. For example, as a patient's health condition deteriorates near end-of-life, increase in symptom burden and decline in functional status may lead to higher level of utilization of both palliative and acute care.[10] If palliative care is analysed as the exposure and ED visits as the outcome event in time-to-event analysis, the hazard ratio (HR) estimate would be biased upward due to the confounding.[11,12] The deterioration in health condition also signifies the beginning of the final stretch of lifespan.[11,13] This time-varying confounding cannot be removed by adjustment or matching for time-constant covariates. Moreover, data capture of time-varying confounders such as palliative care needs is not commonly available in real-world data and patients may be too ill to respond to survey assessment.

The Cox model and its extensions are major candidates in time-to-event analysis. The Cox model is for analysis of a single episode of an outcome event; the Andersen-Gill (AG) model is one of its extensions suitable for analysis of multiple episodes of non-terminal outcomes such as ED visits.[14,15] We refer to them as Cox-type models collectively. They allow researchers to choose the time-scale according to context, though time-on-study (TOS, or $t$ in statistical notation) may be employed without deliberation.[16,17] The influence of the chosen time-scale variable and its correlates on the outcome event rate is cancelled out in the partial likelihood of the models. The potential time-varying confounding related to the time-scale is therefore non-parametrically adjusted for. It is recommended that the time dimension that has the strongest relationship with the outcome should be chosen as the time-scale.[16] There is strong evidence that age is a better choice for time-scale when it has strong impact on the outcomes.[17,18] In studies of infectious diseases, using calendar time as time-scale has the advantage of controlling the confounding by seasonality or changing incidence.[16,19] In studies of advanced illness, time-to-death (TTD) is a strong correlate of many outcomes such as symptom severity, functional decline and healthcare utilization.[10,11] This motivated our research.

We propose "reverse time-to-death" (rTTD, or $t^*$ in statistical notation) as the time-scale in Cox-type models in studies of exposures that tend to occur near end-of-life, where $t^* = \text{TTD}_{\max} - \text{TTD}$, and $\text{TTD}_{\max}$ is the maximum TTD value in a study. This serves as a proxy of observed or unobserved time-varying confounders that intensify as patients approach end-of-life and mitigates their confounding effects. This analytic strategy is straight-forward in studies of decedents, which is common in advanced illness and palliative care research. We also consider a procedure to estimate expected time-to-death among patients who were alive at the end of the study period and use it in the rTTD analysis. In our empirical case study, we hypothesised that: (a) Analysis of ED visits in relation to palliative care uptake using TOS as time-scale would give larger HR than rTTD, representing a higher level of uncontrolled confounding in the former analysis. (b) Analysis with TOS as time-scale would show larger changes in HR estimates when observed time-varying covariates are omitted from statistical adjustment than analysis with rTTD, representing the ability of rTTD time-scale to control unobserved confounding that intensifies in relation to time-to-death.

**METHODS**

**Reverse time-to-death as time-scale**

Suppose the first three of N hypothetical study participants died at two, three and four years after study enrolment, and the fourth participant was followed for one year and then dropped out (censored). Their follow-up times are shown in Figure 1a using TOS as the time-scale. For visual clarity, we plot only the first four of N participants in this figure.

[Figure 1 about here]

All participants entered the study at time *t*=0. Furthermore, participant 2 had an outcome event at *t*=2.5 years. This event time contributes to the Cox-type model's partial likelihood (*PL*) through:[16]

$$PL(t = 2.5) = \frac{\lambda(t = 2.5)exp(\boldsymbol{\beta X}_2)}{\lambda(t = 2.5)exp(\boldsymbol{\beta X}_2) + \lambda(t = 2.5)exp(\boldsymbol{\beta X}_3) + \cdots} = \frac{exp(\boldsymbol{\beta X}_2)}{exp(\boldsymbol{\beta X}_2) + exp(\boldsymbol{\beta X}_3) + \cdots}$$

where $\lambda(t)$ is the baseline hazard at time *t*, $\boldsymbol{X}_i$ is a column vector of covariate values of participant *i*, $\boldsymbol{\beta}$ is a row vector of log(HR) to be estimated, and "…" represents the contributions to $PL(t = 2.5)$ by the other participants who were at risk at $t=2.5$ (not shown in figure). In the above annotation, "baseline" means people whose observed covariate values are all zero. Cox-type models estimate $\boldsymbol{\beta}$ by maximizing the logarithm of the model $PL$, which is the product of all $PL(t)$ arising from the event times.

The intersections of the dotted line and solid lines in Figure 1a indicate "at risk" persons at the event time. Participants 1 and 4 were not included in the denominator of $PL(t = 2.5)$ because they were not at risk at this time. The baseline hazard at $t=2.5$, $\lambda(t = 2.5)$, is cancelled out in $PL(t = 2.5)$ as it appears in both the numerator and denominator.[16,18] In words, when participant 2 had an event at 2.5 years after enrolment, his/her hazard is compared with that of participant 3 and other people (if any) who were also at risk at 2.5 years after enrolment. Therefore, the influence of TOS and its correlates is cancelled out. However, at the time of the event participant 2 was only half a year from death, whereas participant 3 still had 1.5 years to go. They were not comparable in terms of TTD and time-varying covariates that change sharply in relation to TTD.

Suppose we used the method to be discussed in the next section to estimate the survival time for participants whose survival time was censored, and the estimated survival time for participant 4 was two years. Figure 1b right-aligns the follow-up time. Time from censoring to estimated time of death was indicated by a dashed line. We call this time-scale reverse time-to-death (rTTD), defined as:

$$t^* = \text{TTD}_{\text{max}} - \text{TTD}$$

where $\text{TTD}_{\text{max}}$ is the maximum time-to-death no matter if the values are observed or estimated. Suppose participant 3 had the longest time-to-death, so $\text{TTD}_{\text{max}} = 4$.

Entry to the study is now staggered. Participant *i* enters the study at $\text{TTD}_{\text{max}} - \text{TTD}_i$, where $\text{TTD}_i$ was the TTD of participant *i* at enrolment. Statistical software like Stata and R allow staggered entry. We used Stata's *stcox* program.[20]

Participant 2 now has an outcome event at $t^* = 3.5$. This event time contributes to the *PL* through:

$$PL(t^* = 3.5) = \frac{\lambda(t^* = 3.5)exp(\boldsymbol{\beta X_2})}{\lambda(t^* = 3.5)exp(\boldsymbol{\beta X_1}) + \lambda(t^* = 3.5)exp(\boldsymbol{\beta X_2}) + \lambda(t^* = 3.5)exp(\boldsymbol{\beta X_3}) + \cdots}$$

$$= \frac{exp(\boldsymbol{\beta X_2})}{exp(\boldsymbol{\beta X_1}) + exp(\boldsymbol{\beta X_2}) + exp(\boldsymbol{\beta X_3}) + \cdots}$$

Participant 1 is now included in the denominator of $PL(t^* = 3.5)$ as s/he was at risk at $t^* = 3.5$. Participant 4 is not included because s/he had left the study at $t^* = 3$. However, if there were participants who had outcome events between $t^* = 2$ and 3, participant 4 would be in the denominator of $PL(t^*)$ at these event times. The hazard at this time-to-death, $\lambda(t^* = 3.5)$, is cancelled out in the *PL*. Therefore, the impact of time-to-death and its correlates on the outcome is eliminated.

In short, when participant 2 had an event half a year before death, his/her hazard of the outcome event is compared with that of people who were also at half a year before death. Using rTTD as time-scale helps to compare like with like in terms of time-to-death and its associated time-varying covariate values.

**Estimation of time-to-death**

Studies of decedents are common in research on advanced illness and palliative care, e.g. the population-based Ontario and Belgium decedent cohorts.[21,22] The rTTD method is straight-forward in this case without the need for estimation of time-to-death.

The Buckley-James (BJ) method is a distribution-free method for estimation of expected survival time for censored observations given a dataset that includes both censored and observed survival times and observed covariates.[23,24] It is implemented in statistical software such as Stata and R. We used Stata's *buckley* program.[25] The robustness may be low if a large proportion of observations is censored. Heller and Simonoff recommended the use of BJ method when censoring proportion is smaller than 40%.[26] Considering the paucity of model diagnostics for the method, Stare et al. recommended a more stringent criterion of <20%.[24] The BJ method is suitable in studies of advanced illness where few patients survived despite its limitation in broader applications.

**Case study: Healthcare utilization in stage IV cancer patients**

The Cost of Medical Care of Patients with Advanced Serious Illness in Singapore (COMPASS) is a prospective cohort study of 600 adult patients (age ≥21) with stage IV solid cancer, recruited between 2016 and 2018 from National Cancer Centre Singapore (NCCS) and National University Hospital System. Details of the study protocol have been published.[27] Findings on ED visits and other acute healthcare utilization in the last month of life has also been published.[28] Briefly, consented patients were interviewed every 3 months until death or 60 months post enrolment, whichever earlier. They also provided consent for access to their electronic health records (EHR) held by their healthcare providers, including acute care hospitals and community-based palliative care providers. The EHR data covered till 31 December 2021. The study is approved by SingHealth Centralized Institutional Review Board (2015-2781) and National University of Singapore Institutional Review Board (S-20-155).

For purpose of illustration of the use of rTTD time-scale, we analysed ED visits in relation to community-based palliative care (PC for brevity), including home and day care and regardless of frequency/duration of utilization. In the present context, PC is a time-varying exposure variable.[16] For example, if a patient started using PC in the mid-point between study enrolment and death, the first half of the person-time will be classified as unexposed and the second half as exposed.

Time-constant covariates included age at enrolment, gender, type of cancer, MediFund status (an indicator of financial difficulty) and education. Time-varying covariates from 3-monthly interviews include the Physical Well-being (PWB) and Functional Well-being (FWB) scores of Functional Assessment of Cancer Therapy – General (FACT-G), which are known predictors of cancer survival.[29,30] Missing PWB and FWB values (16 and 17, respectively, out of totally 5499 survey questionnaires) were handled by last-observation-carried-forward.

**Statistical analysis**

Since one person could have multiple ED visits, we used the AG model for the analysis and robust standard error for cluster data for inference.[14,15] Bootstrapping was used to estimate confidence intervals for difference in HR estimates between different models, with persons as resampling units and 1000 replicates. For graphical presentation of hazard functions, we used kernel smoothing with boundary-bias correction.[20]

We began with analysis of patients who were deceased by end of 2021. Then, for the full cohort analysis, we used the aforementioned time-constant covariates and baseline PWB and FWB scores as predictors in the BJ method to estimate survival time for patients who were alive at the end of 2021. After 2021, manual review of medical records at NCCS found the date of death of 32 patients who died in 2022 or 2023. We used these 32 records to evaluate the accuracy of the BJ analysis, in which they were kept censored at the end of 2021, by comparing their observed and BJ-estimated survival times.

**RESULTS**

Two of 600 patients were excluded from analysis due to missing covariate data. By the end of 2021, 429 of 598 patients had died. Among the decedents, 207 patients had ever used PC during the study period. There were 885 ED visits and 651.5 person-years of observation (Table 1). The incidence rate (number of ED visits per person-year) was lower among patients who did not use PC than patients who did (1.19 versus 1.55). Among the latter group, incidence rate was lower before their starting PC than after (1.06 versus 3.10).

[Table 1 about here]

Figures 2a to 2d show the smoothed hazard estimates of ED visits and smoothed mean PWB scores among the decedents. Using TOS time-scale, there was a wide gap in hazard of ED visits between person-time exposed and unexposed to PC and the hazard was roughly stable over time except at the tail ends (Figure 2a). In contrast, using rTTD time-scale reveals that the hazard increased as patients approached end-of-life and that person-time on PC had higher hazard mainly because this was nearer end-of-life (Figure 2b). Comparing the estimates at the same time-to-death, the difference in hazard between person-time on and not on PC was much smaller than in Figure 2a.

[Figure 2 about here]

From the TOS perspective (Figure 2c), there was a gap of about three to four points in mean PWB score between PC and no PC. Using the rTTD perspective reveals that PWB scores were lower in person-time on PC mainly because they were nearer end-of-life (Figure 2d). Given the same time-to-death, the difference in PWB between the two curves was only about one point. Statistical adjustment for PWB would therefore make a substantial impact on the HR estimates in analysis of ED utilization based on TOS but not rTTD.

Among the decedents, without controlling for any covariates, the HRs were 2.71 and 1.31 with TOS and rTTD as time-scale, respectively (Table 2). The difference in HR was 1.40 (95% CI 0.97 to 1.82). The finding was consistent with our first hypothesis that using TOS as time-scale would give a higher HR estimate than rTTD. Adjustment for time-constant covariates made little difference to the HR estimates. This trivial change-in-estimate should not be taken as evidence of no time-constant confounding because unobserved time-constant covariates may be dominant.

[Table 2 about here]

The results were also consistent with our second hypothesis that omission of adjustment for time-varying covariates would make larger change-in-estimate in analysis using TOS than rTTD. The adjustment led to difference in HR between 2.72 and 2.34 using TOS (change-in-estimate=0.38; 95% CI 0.15 to 0.60). In contrast, the adjustment only made a trivial difference in HR between 1.32 and 1.28 using rTTD (change-in-estimate=0.04; 95% CI -0.02 to 0.11).

In analysis of the full cohort, time-to-death was estimated using the BJ method for 169 patients who were alive at the end of 2021. They were mostly non-PC users who had low ED utilization, leading to lower incidence of ED visits in the non-PC users (0.58 per person-year, Table 1) and higher HRs in the full cohort than decedents (Table 2). However, the pattern of HR estimates between the two time-scales was similar to that in the decedent analysis.

The mean observed and estimated time-to-death of the set of 32 observations earmarked for evaluation of the BJ method were 5.53 and 6.29 years, respectively. The mean absolute difference was 1.0 year. Thus, the right-alignment of the survivors' follow-up times as illustrated in Figure 1b might have been somewhat inaccurate. This could generate residual confounding. This may explain why adjustment for time-varying confounders led to a larger change in HR (2.05 vs 1.85, change-in-estimate=0.20; 95% CI 0.05 to 0.35) in the rTTD analysis in the full cohort than in decedents only. Nevertheless, it was still much less than the change-in-estimate between 4.20 and 3.09 using TOS (change-in-estimate=1.11; 95% CI 0.59 to 1.52).

**DISCUSSION**

The choice of time-scale in Cox-type models offers a simple and robust way to control time-varying confounding. A general recommendation is to choose the time dimension that has the strongest relationship with the outcome.[16,18] In advanced illness, the level of many outcomes change sharply near end-of-life, making time-to-death a suitable choice.

Some advanced illness studies involve decedents only, but some also involve patients who were alive at the end-of-study. The Buckley-James method can be used to estimate their survival times. The robustness of the analysis is affected by the proportion of participants with censored survival time. Studies of advanced illness with only a small proportion of censored observations, preferably <20% and at most 40%,[24,26] may consider using this approach.

This study focused on time-varying confounding. It is important to also search for better approaches to handle time-constant confounding. We do not interpret the presented analytic results from COMPASS as an indication of palliative care leading to higher rate of ED visits because unobserved time-constant confounders such as psychosocial factors is still an issue.[31-33] The prior event rate ratio approach that is gaining popularity in biopharmaceutical research is basically a ratio-of-ratio analysis.[34,35] Conceptually this is similar to difference-in-difference analysis that is popular in policy research.[8,9] This method aims to control the impact of observed or unobserved time-constant confounders in time-to-event analysis. Combined use of reverse time-to-death and prior event rate ratio appears promising and is an area for future research.

As compared to other advanced illness, the progression trajectory of heart and lung failures shows more short-term fluctuations.[11] The applicability and benefits of the proposed method in these conditions will need further exploration.

**Table 1**. Incidence rate (number of events / person-years) of emergency department visit, by exposure to palliative care

| Sample | Overall | Non-PC* users | PC* users | | |
|---|---|---|---|---|---|
| | | | All time | Before PC | After PC |
| Decedents (n=429) | 1.36 (885/651.5) | 1.19 (407/343.4) | 1.55 (478/308.1) | 1.06 (247/233.5) | 3.10 (231/74.6) |
| Full cohort (n=598) | 0.79 (1147/1452.9) | 0.58 (664/1140.4) | 1.55 (483/312.6) | 1.05 (248/236.3) | 3.08 (235/76.3) |

* PC: palliative care

**Table 2**. Hazard ratios of emergency department visit in person-time exposed to palliative care versus unexposed, from Andersen-Gill models using different time-scales

| Sample | Time-scale [*] | Covariate adjustment [**] | | | | | |
|---|---|---|---|---|---|---|---|
| | | None | | TCC only | | TCC + TVC | |
| | | HR[†] | 95% CI | HR | 95% CI | HR | 95% CI |
| Decedents | TOS | 2.71 | (2.14, 3.43) | 2.72 | (2.15, 3.44) | 2.34 | (1.85, 2.97) |
| (n=429) | rTTD | 1.31 | (1.04, 1.66) | 1.32 | (1.05, 1.66) | 1.28 | (1.02, 1.60) |
| Full cohort | TOS | 4.27 | (3.38, 5.38) | 4.20 | (3.32, 5.32) | 3.09 | (2.45, 3.90) |
| (n=598) | rTTD | 2.12 | (1.67, 2.69) | 2.05 | (1.61, 2.60) | 1.85 | (1.46, 2.34) |

[*] TOS: time-on-study; rTTD: reverse time-to-death.

[**] TCC: time-constant covariates; TCC+TVC: time-constant and time-varying covariates; see Methods section for covariates included.

[†] HR: hazard ratio.

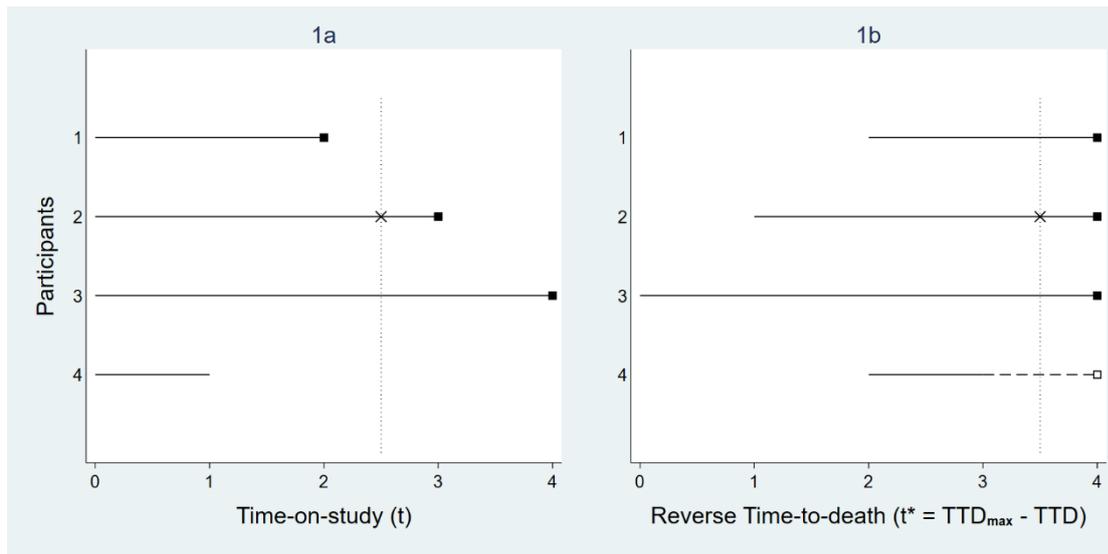

**Figure 1**. Illustration of (a) time-on-study ($t$) and (b) reverse time-to-death ($t^*$) as time-scales in Cox-type models. ■ observed time of death; × observed time of outcome event; □ estimated time of death; solid line: actual follow-up time; dashed line: time between end of actual follow-up and estimated time of death; intersections of dotted line and solid lines indicate "at risk" persons at the event time.

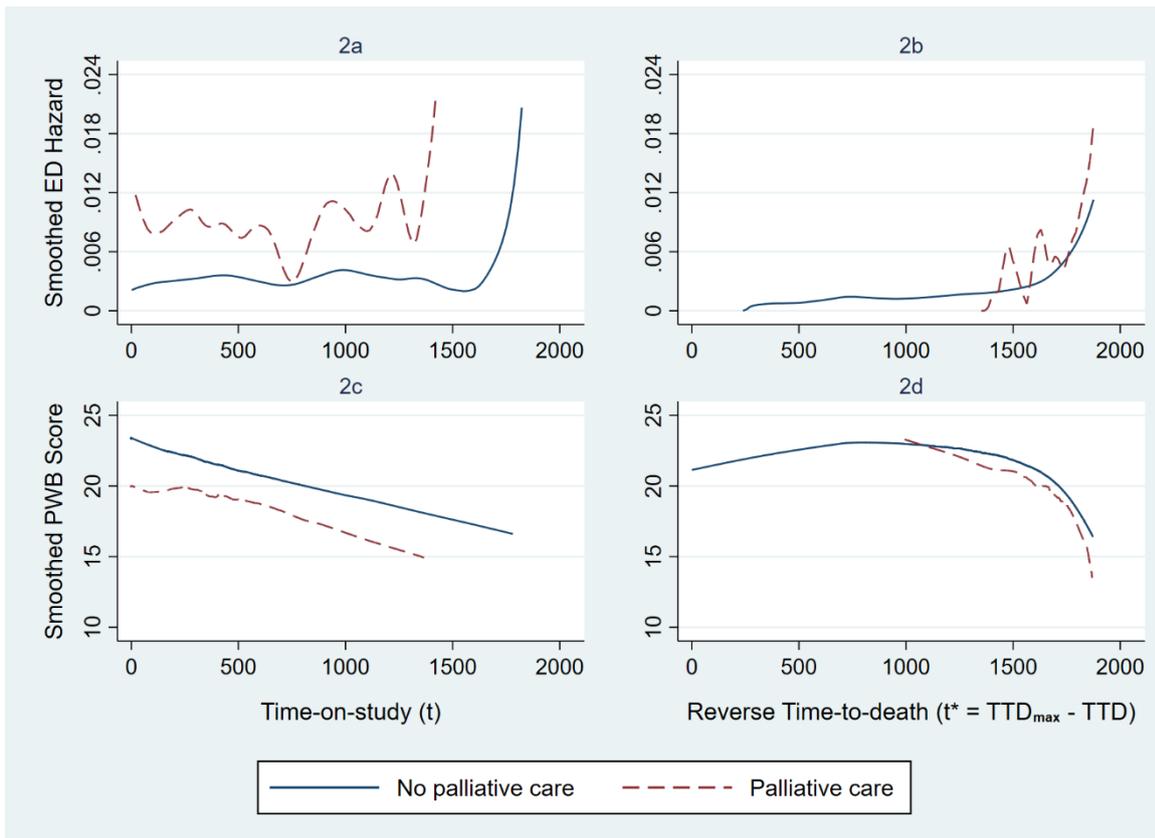

**Figure 2**. Smoothed estimates of hazard of emergency department (ED) visit and smoothed mean Physical Well-being (PWB) score by time-scale and exposure to palliative care; decedents only.

**Competing interest**

None declared.

**Contributors**

YBC is the guarantor. YBC planned the study and wrote and revised the manuscript. XM performed data analysis and revised the manuscript. NL, QZ and GMY planned the study and revised the manuscript. CM and EAF planned the study, collected the data, and revised the manuscript. IC collected and managed the data, managed the project and revised the manuscript. All authors read and approved the final manuscript.

**Funding**

The methodological work was supported by the National Medical Research Council, Singapore (MOH-001487). The COMPASS study was funded by Singapore Millennium Foundation (2015-SMF-0003) and Lien Centre for Palliative Care (LCPC-IN14-0003). Any opinions, findings and conclusions or recommendations expressed in this material are those of the authors and do not reflect the views of Ministry of Health / National Medical Research Council, Singapore, Singapore Millennium Foundation, or Lien Centre for Palliative Care.

**Ethics approval**

The COMPASS study is approved by SingHealth Centralized Institutional Review Board (2015-2781) and National University of Singapore Institutional Review Board (S-20-155).

**Patient consent for publication**

Not applicable.

**Data availability statement**

COMPASS study data are not publicly available due to restrictions on distribution of electronic health record data. Data are however available from the authors upon reasonable request and permission of SingHealth and National University Health Systems and institutional review board approval from the requesting institution.